\documentclass[conference]{IEEEtran}
\usepackage{graphicx} 
\usepackage{cite}
\usepackage{color, soul}

\ifCLASSINFOpdf
\else
\fi
\begin{document}
\title{Programmable Metasurfaces:\\ State of the Art and Prospects}
\author{
	\IEEEauthorblockN{
		Fu Liu\IEEEauthorrefmark{1},
		Alexandros Pitilakis\IEEEauthorrefmark{2}\IEEEauthorrefmark{3}, 
		Mohammad Sajjad Mirmoosa\IEEEauthorrefmark{1}, 
		Odysseas Tsilipakos\IEEEauthorrefmark{2}, \\
		Xuchen Wang\IEEEauthorrefmark{1}, 
		Anna C. Tasolamprou\IEEEauthorrefmark{2}, 
		Sergi Abadal\IEEEauthorrefmark{4}, 
		Albert Cabellos-Aparicio\IEEEauthorrefmark{4}, \\
		Eduard Alarc\'{o}n\IEEEauthorrefmark{4}, 
		Christos Liaskos\IEEEauthorrefmark{2}, 
		Nikolaos V. Kantartzis\IEEEauthorrefmark{2}\IEEEauthorrefmark{3},\\
		Maria Kafesaki\IEEEauthorrefmark{2}\IEEEauthorrefmark{5}, 
		Eleftherios N. Economou\IEEEauthorrefmark{2}, 
		Costas M. Soukoulis\IEEEauthorrefmark{2}\IEEEauthorrefmark{6},
		Sergei Tretyakov\IEEEauthorrefmark{1}, 
	}
	\IEEEauthorblockA{\IEEEauthorrefmark{1}Department of Electronics and 
		Nanoengineering, Aalto University, P.O. Box 15500, Espoo, Finland}
	\IEEEauthorblockA{\IEEEauthorrefmark{2}Foundation for Research and 
		Technology Hellas, 71110, Heraklion, Crete, Greece}
	\IEEEauthorblockA{\IEEEauthorrefmark{3}Department of Electrical and 
		Computer Engineering, Aristotle University of Thessaloniki, 
		Thessaloniki, Greece}
	\IEEEauthorblockA{\IEEEauthorrefmark{4}NaNoNetworking Center in 
		Catalonia (N3Cat), Universitat Polit\`{e}cnica de Catalunya, 
		Barcelona, Spain}
	\IEEEauthorblockA{\IEEEauthorrefmark{5} Department of Materials 
		Science and Technology, University of Crete, 71003, Heraklion, Crete, 
		Greece}
	\IEEEauthorblockA{\IEEEauthorrefmark{6} Ames Laboratory and Department 
		of Physics and Astronomy, Iowa State University, Ames, Iowa 50011, USA\\
		Email: fu.liu@aalto.fi
	}
}

\maketitle

\begin{abstract}
Metasurfaces, ultrathin and planar electromagnetic devices with sub-wavelength unit cells, have recently attracted enormous attention for their powerful control over electromagnetic waves, from microwave to visible range. With tunability added to the unit cells, the programmable metasurfaces enable us to benefit from multiple unique functionalities controlled by external stimuli. In this review paper, we will discuss the recent progress in the field of programmable metasurfaces and elaborate on different approaches to realize them, with the tunability from global aspects, to local aspects, and to software-defined metasurfaces.
\end{abstract}
\IEEEpeerreviewmaketitle

\section{Introduction}
Metamaterials and metasurfaces are artificial bulk and planar materials with subwavelength structural inclusions. Benefiting from the tailored unit cell structures, metasurfaces have been showing powerful abilities in achieving versatile functionalities, such as perfect absorption, anomalous reflection, focusing, imaging, to mention a few~\cite{GLYBOVSKI20161,chen2016}. In the early stages of metamaterials and metasurfaces research, once the unit cell is designed, its function is fixed, for example, an absorber works at a certain frequency where the input impedance is matched to the free space. However, if we want to change the working frequency or even the functionality, re-design and re-fabrication processes are inevitable due to the structural nature of the unit cells. In fact, the properties of the metamaterials and metasurfaces can be adjusted by adding \textquotedblleft tuning\textquotedblright~capability in the unit cells~\cite{shadrivov2017,Zheludev2016,Turpin2014}. Then their electromagnetic wave behavior can be tuned externally by modifying the stimuli. Moreover, such tuning can be controlled by a computer program. From this point of view, such metasurfaces are programmable and they provide more opportunities in achieving dynamical wave applications, without the re-fabrication process. 

There are many tuning mechanisms for realizing programmable metasurfaces. For instance, the change of the properties of the substrate or structural material, e.g., the permittivity of the liquid crystal can be changed under different gate voltages~\cite{Zhang2010}, can result in a shift of the resonance frequency. Changing the \textquotedblleft on\textquotedblright~and \textquotedblleft off\textquotedblright~state of a diode will switch the metasurface performance from perfect absorption to reflection~\cite{Zhu2010diode}. In this paper, we will review different tuning mechanisms in achieving programmable metasurfaces, from global tuning at the metasurface level to local tuning at the unit cell level. Finally we will direct our discussion to the software-defined metasurfaces which have the ability to tune the properties of each unit cell independently.

\section{Global tuning at the metasurface level}
\begin{figure*}[ht]
	\centering
	\includegraphics[width=410pt]{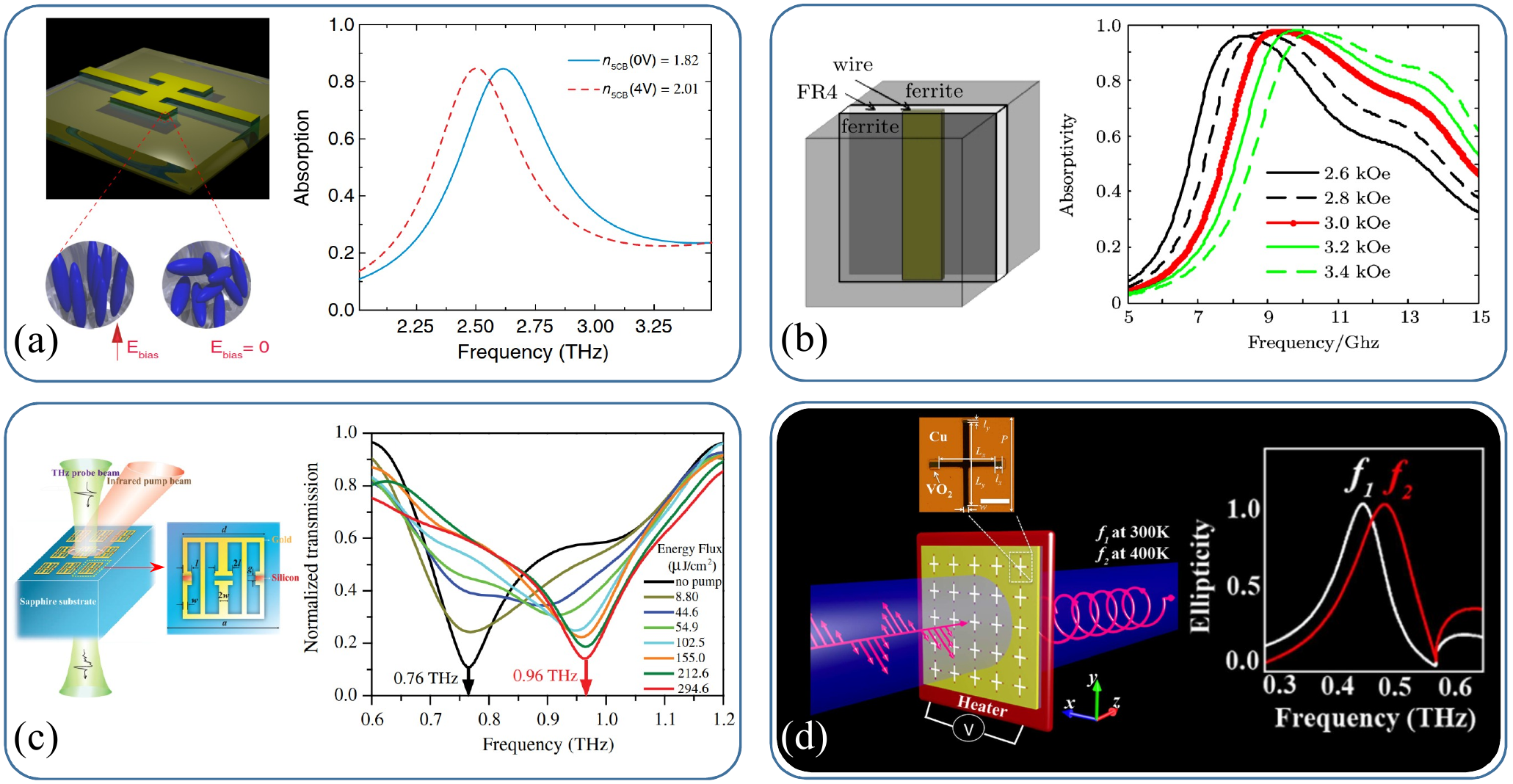}
	\vspace{-0.3cm}
	\caption{\label{figglobaltuning} 
		Tunable metasurfaces with stimuli-sensitive materials. 
		(a) Electric-sensitive liquid crystal-based tunable metasurface absorber~\cite{Shrekenhamer2013}.
		(b) Magnetic-sensitive ferrite-based tunable metasurface absorber~\cite{Yang:2012}.
		(c) Light-sensitive semiconductor-based tunable metasurface~\cite{Shen:2011}.
		(d) Thermal-sensitive VO$_2$-based tunable quarter-wave plate~\cite{Wang:2015}.
	}
	\vspace{-0.3cm}
\end{figure*}

We can make the metasurface programmable by introducing stimuli-responsive materials, which are capable of undergoing relatively large and rapid changes in their physical properties in response to external ambient stimuli~\cite{Sieklucka:2016}. In this case, when the ambient conditions, such as temperature, pressure, humidity, electric/magnetic field, or light are altered, then the material properties will be tuned accordingly and therefore induce a change of the metasurface functionality. As the ambient stimuli apply to the whole metasurface level, we refer to this approach as global tuning. As the ambient pressure and humidity is hard to change, we will focus on the tunability by electric, magnetic, light, and temperature stimuli, with some examples shown in Fig.~\ref{figglobaltuning}. 

\subsection{Electric tuning}
There are many electrically sensitive materials, among which nematic liquid crystals (LCs) are very famous due to the massive development of the optical display technology. Nematic LC molecules respond to the bias electric field by rearranging their orientation and thus achieving voltage dependent birefringence. Due to their liquid nature, they can be infiltrated into various metasurface structures providing large refractive index modulation for operation in the microwave, terahertz and optical regime~\cite{Zhang2010,Zografopoulos2015,Decker20138879,Ou2013252}. To date, nematic LCs have been used in various pronounced metasurface applications such as tunable absorbers, as shown in Fig.~\ref{figglobaltuning}(a) \cite{Shrekenhamer2013}, and cloaks with possibilities of real-time control of invisibility \cite{Pawlik20121847}. Due to their large anisotropy, LCs have also been employed in hyperbolic metasurface configurations~\cite{Cao20161753_2016}.

Another famous electrically sensitive material is graphene, whose Fermi level can be modified by the external electrostatic field. For example, in \cite{ju2011graphene}, researchers have demonstrated tunable plasmon resonances in graphene microribbon arrays on a silicon/silicon dioxide substrate. However, the weak interaction between graphene and light, which is due to the poor obtainable mobility (especially for processed graphene), hinders the tunable functionalities. In fact, from the circuit perspective, the weak interaction comes from the huge impedance mismatch between graphene and the surrounding materials. While the low-quality graphene has very large surface impedance in the terahertz range, the characteristic impedance of the surrounding materials is comparably small. 
To circumvent this low tunability, metallic blocks are introduced on the graphene to largely reduce the effective surface impedance. In this way, the graphene-metal hybrid metasurfaces have shown efficient tunability on amplitude, phase, and resonance frequency~\cite{lee2012switching,jadidi2015tunable,dabidian2016experimental,kim2017electronically,Wang2017}. For example, a recent work in \cite{kim2017electronically} shows that a graphene-metal strip structure can demonstrate strong tunable absorption (with absorbance modulation efficiency of 96\%) when the Fermi level is tuned from 0.26~eV to~0.57 eV.

\begin{figure*}[ht]
	\centering
	\includegraphics[width=480pt]{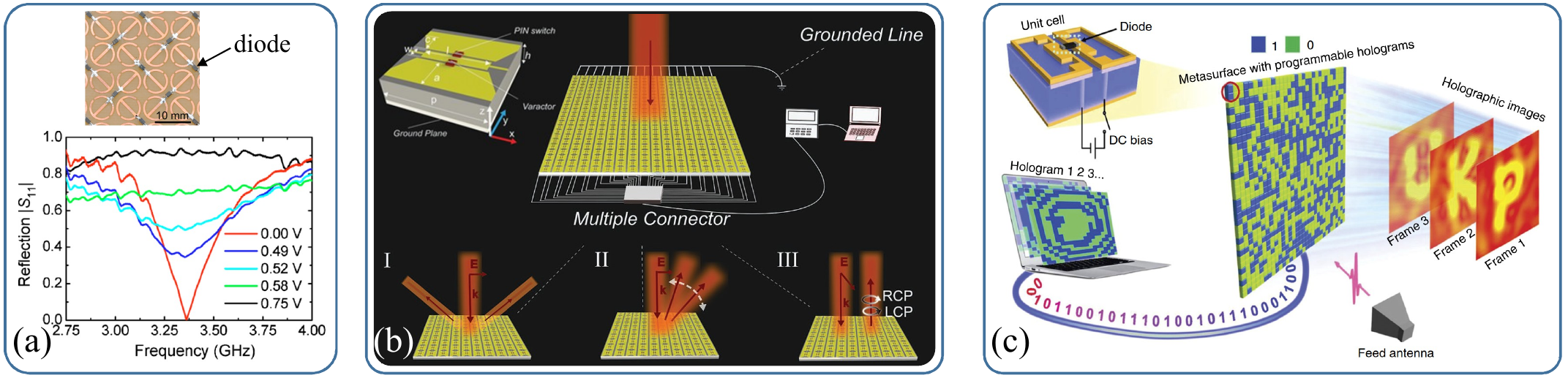}
	\vspace{-0.3cm}
	\caption{\label{figvaractorloaded} 
		Varactor-loaded tunable metasurfaces with incrasing complexity and ability. 
		(a) A diode switches the functionality from total absorption to total reflection.~\cite{Zhu2010diode}.
		(b) Column-controlled varactors enable a tunable metasurface to perform multiple functions: splitting, steering, and polarization conversion~\cite{Huang2017}.
		(c) Patterned coding-metasurface enables dynamic hologram creation~\cite{Li2017hologram}.
	}
	\vspace{-0.3cm}
\end{figure*}

\subsection{Magnetic tuning}
Magnetically tunable structures are attractive thanks to their instantaneous response to external contactless magnetic stimuli. The magnetic field-guided self-assembly of colloidal particles is one of the leading fabrication methods: The induced dipole-dipole interactions between these particles can be accurately controlled (attractive or repulsive), so that by properly adjusting the controlling magnetic field profile their assembly process can be tuned, from the movement of a single nano-sized object to a globally synchronized motion. Applications include adaptive microplate arrays, i.e., magnetically responsive nanostructures for tuned optics~\cite{Liu:2017}, hybrid devices~\cite{Pandey:2012} and generation of resonant-assisted tunneling phenomena~\cite{Bourquin:2014}. Magnetically responsive structures have been exploited also in tunable-resonance metamaterials: ferrite-wire metamaterials for tunable negative index~\cite{He:2006}, ferrite-rod structures with different saturation magnetization for wide-band tunable microwave filters~\cite{Bi:2015}, and tunable broadband absorbers consisting of ferrite slabs and copper wires~\cite{Yang:2012}, as shown in Fig.~\ref{figglobaltuning}(b).

\subsection{Light tuning}
Photoconductive semiconductor materials, such as Si and GaAs, have been employed for implementing tunable metasurfaces by tuning their conductivity through carrier photoexcitation with an infrared pump beam. Several implementations with split-ring-resonator-based metasurfaces operating in the THz have been proposed. In \cite{Padilla:2006,Chatzakis:2012}, the semiconductor (GaAs) is included as the substrate material. Alternatively, the semiconductor (Si) can be incorporated as a section in the resonant structure~\cite{Chen:2008,Shen:2011}. In this case, carrier photoexcitation effectively modifies the resonator geometry. This can lead to a redshift of the fundamental resonance~\cite{Chen:2008} or trigger a transition to a different, blue-shifted resonance~\cite{Shen:2011}, as shown in Fig.~\ref{figglobaltuning}(c). Semiconductor inclusions have been also utilized in chiral metasurfaces for switching their handedness~\cite{Zhang:2012} or tuning the ellipticity and optical activity~\cite{Kenanakis:2014}.

\subsection{Thermal tuning}
Heat is another ambient stimulus for tunable metasurfaces. Metamolecules or substrates whose macroscopic parameters are sensitive to temperature variations can lead to the change of the electromagnetic response of the metasurfaces. In this context, phase change materials (PCMs) with temperature-dependent permittivity are promising candidates~\cite{Driscoll:2008,Driscoll:2009,Seo:2010,Kim:2015,Wang:2015,Shin:2015,Wang:2016}. For instance, vanadium dioxide, as one of the well-known PCMs, behaves as an insulator at room temperature and transits to a metal state at higher temperatures due to the enhancement of the free carriers concentration. It has been employed in tunable metasurfaces to  shift the resonance frequency, modulate the transmission amplitude~\cite{Driscoll:2008,Driscoll:2009}, or switch  the polarization of the transmitted wave at two different frequencies~\cite{Wang:2015}, as shown in Fig.~\ref{figglobaltuning}(d). A structure which consists of two materials with different thermal expansion coefficients is also a plausible solution. In \cite{ou:2011}, a metasurface made of such a structure can have different transmission amplitude when the temperature varies. Moreover, superconductors can be also employed for tuning the transmission resonance and its strength since the conductivity of superconductors is temperature dependent~\cite{Chen:2010}. 

\section{Local tuning at the unit-cell level}
While the global tuning applies to the whole metasurface, local tuning provides a possibility to tune the properties of each unit cell locally. In this way, we can achieve more tunable functions, such as steering, imaging and holography. For local tuning, we can  use the same stimuli-sensitive materials. But in this case we need to apply the stimuli locally in each unit cell. For example, we need a heater/cooler in each unit to effectively control the temperature, a coil (capacitor) in each unit to change the local magnetic (electric) field, and a LED in each unit to apply different illumination~\cite{Shadrivov2012lighttune}. These, however, are usually impractical solutions due to additive large objects compared to the unit cell size even in the GHz-range  metasurfaces. Probably the best option is to use voltage-driven elements, for example diodes and varactors, which have comparably small sizes. Fig.~\ref{figvaractorloaded} lists some examples with different levels of complexity.

\subsection{Switch diode}
A switch diode, which has two states \textquotedblleft on\textquotedblright~and \textquotedblleft off\textquotedblright, enables dual-function metasurfaces in the GHz band, if we apply the same control voltage on the diodes. For example, Fig.~\ref{figvaractorloaded}(a) shows one design in which total absorption and total reflection can be switched by changing the state of the diodes~\cite{Zhu2010diode}. In this case, the diode changes the input impedance of the metasurface to match (total absorption) or mismatch (total reflection) with the free space impedance. In another design~\cite{Tao2017diode}, both the polarization and scattering properties can be modified. While the incident linearly polarized wave is reflected to the same polarization when the diodes are on, the same incident wave will be transmitted with perfect polarization rotation when the diodes are switched off.

\subsection{Continuous tuning varactors}
Actually, the varactor diodes (capacitors) can be adjusted in a continuous way~\cite{Zhao2013NJP}. In the simplest scenario, all varactors are controlled by one and the same voltage, which effectively gives frequency tunability to the functionality the metasurface is designed for~\cite{Mias2007TAP,Burokur2010APL,Dincer2015JEWA,Masud2012TEMC}. The most widely investigated functionality is tunable perfect absorption, where a change in the reverse biasing voltages of the varactors shifts the spectrum of the perfect absorption resonance~\cite{Ma2016Optik,Zhu2015EPL,Luo2016APL,Kim2016AO}. Typical values for the (reverse) voltage range are 0-20 V corresponding to an equivalent capacitance range of a few pico-Farad, e.g. 0.5-3.5 pF, for commercially available diodes that are compact enough to be integrated on these GHz-band metasurfaces. The corresponding frequency tunability accessible by this capacitance range is in the order of a couple of GHz, e.g. 4-6 GHz. Similarly, microelectromechanical systems (MEMS), in which the capacitances are tuned by the piezoelectric effect, provide another option for continuously tunable high impedance metasurfaces~\cite{Chicherin2006MEMS}.

\subsection{Collective tuning varactors}
Moving one step towards more elaborate functionality, the locally-applied continuous tuning voltage is allowed to be different for each unit cell of the metasurface. For example, we can apply a voltage profile on the varactors in one direction while keeping the voltage unchanged in the other direction, as shown in Fig.~\ref{figvaractorloaded}(b). In this way, a specific one dimensional phase profile can be actively \textquotedblleft imprinted\textquotedblright~on the metasurface. This enables us to achieve a range of tunable applications, such as tunable reflection (steering)~\cite{Sievenpiper2003TAP,Xu2016SREP,Chicherin2008MEMS,Huang2017}, beam splitting~\cite{Huang2017}, and even writing a letter by dynamically changing the focal point with a tunable \textquotedblleft Huygens' metasurface\textquotedblright~\cite{Chen2017}. Evidently, the price to pay for this enhanced functionality is the increased complexity in the electrical network that controls the voltages biasing the varactor diodes of the metasurface.

On the other hand, instead of modulating the unit cell properties only in one direction, we can dynamically program the metasurface into full two dimensional patterns. In this case, all individual varactors can be controlled independently and therefore we can obtain a huge number of available patterns, which have enabled researchers to demonstrate dynamical hologram~\cite{Li2017hologram}, as shown in Fig.~\ref{figvaractorloaded}(c). Of course, the electrical network of such programmable metasurface is very complex and we need to feed 20$\times$20 group of diodes. A simpler way is to use two diodes in each unit cell and connect one in column and one in row. By doing so only two sets of control voltages are required, while the programmable metasurface can still give promising functionality, for example, single-sensor imaging by utilizing many available configuration patterns~\cite{Li2016imaging}.

\section{Toward software-defined metasurfaces}
\begin{figure}
	\centering
	\includegraphics[width=210pt]{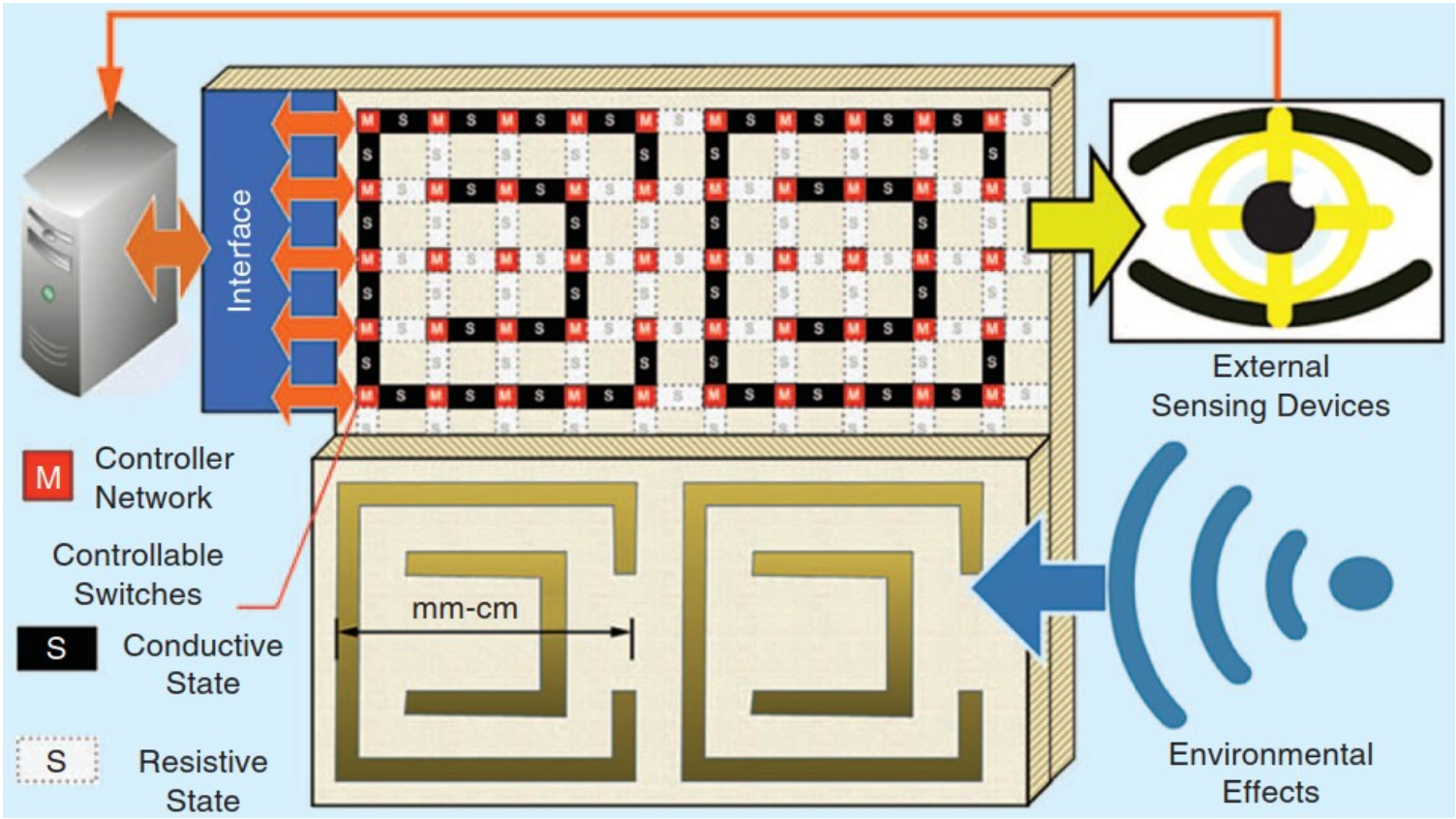}
	\vspace{-0.2cm}
	\caption{\label{figSDM} Software-defined programmable metasurface enables structural reconfiguration at the unit cell level~\cite{SDM}.
	}
	\vspace{-0.2cm}
\end{figure}

A more elaborate way in making tunable metasurfaces is to have the tunability in the change of the unit cell structural configurations. Differently from those approaches which mechanically change the structure shape by external stimuli, such as piezoelectric materials~\cite{Chicherin2006MEMS,Chicherin2008MEMS} and thermal-sensitive materials~\cite{ou:2011}, this approach reconnects  parts of the unit cell to achieve different structural configurations, as proposed in \cite{SDM} and shown in Fig.~\ref{figSDM}. By controlling the connectivity at different locations, reconfiguration of the structure is obtained, thus providing multiple and tunable functionalities. At this point, one challenge resides in the development of a platform that allows users to easily characterize and reconfigure the metasurface. A few recent works~\cite{SDM,SDM2} aim to achieve this goal using a software-defined approach, where there is a clear separation between electromagnetic functions offered by the metasurface at the macroscopic level and the unit cell configuration that yields them. 

The realization of a software-defined metasurface requires a hardware system that applies the software primitives and effectively reconfigures the metasurface. To this end, some researchers propose to integrate a network of tiny controllers within the metasurface structure and to wirelessly interface it with an external entity~\cite{SDM,SDM2}. Each controller is capable of interpreting global instructions and of acting locally by, for instance, tuning its corresponding varactors to achieve the desired impedance configuration. The main challenges here are, firstly, to develop an ultra-low cost network of controllers and, secondly, to co-integrate it with the metasurface within a single structure.

In the software-defined platform, the metasurface profile, which refers to the systematic, measurements-based registration of functions offered by a given metasurface, constitutes the linchpin between academic knowledge and real-world applications. The information contained in the profile allows software developers and engineers to design systems that contain the electromagnetic behavior of objects into their control loops, without required knowledge of the underlying physics. This evolution comes as a timely extension of the concept of Internet-of-Things (IoT)~\cite{Mihovska.2018}, which constitutes a robust, complete hardware platform for connecting anything-to-anything, under a considerable range of conditions and use-cases. Novel IoT products are being released almost daily, at a trend that is expected to yield 20-30 billion connected IoT devices by 2020~\cite{Nordrum.2016}. Software-defined metasurfaces can give the IoT concept a new application field over the electromagnetic behavior of objects. Coupled with the efforts seeking to provide control over mechanical properties~\cite{claytronics}, IoT can extend to Internet-of-Materials (IoM), offering unprecedented capabilities. 

\section{Conclusion}
Tunable metasurfaces benefiting from both global and local tuning have led us to a greater control of electromagnetic waves by artificial skins. The capacity to host multiple functionalities concurrently, or switch between them, opens a door to a myriad of applications. We envisage that the software-defined metasurfaces have the potential to automatically adapt to environmental changes.

\section*{Acknowledgment}
This work was supported by the European Union’s Horizon 2020 research and innovation programme-Future Emerging Topics (FETOPEN) under grant agreement No 736876.

\bibliographystyle{IEEEtran}
\bibliography{IEEEabrv,progmeta}

\end{document}